# Martensitic transformation in V$_3$Si single crystal: $^{51}$V NMR evidence for coexistence of cubic and tetragonal phases

A. A. Gapud[*,1], S. K. Ramakrishna[2], E. L. Green[2], and A. P. Reyes[2]

[1]University of South Alabama, Department of Physics, 411 N. University Blvd. MPSB 115, Mobile, AL 36688

[2]The National High Magnetic Field Laboratory, 1800 E. Paul Dirac Drive, Tallahassee, FL 32310

## Abstract

The Martensitic transformation (MT) in A15 binary-alloy superconductor V$_3$Si, though studied extensively, has not yet been conclusively linked with a transition to superconductivity. Previous NMR studies have mainly been on powder samples and with little emphasis on temperature dependence during the transformation. Here we study a high-quality single crystal, where quadrupolar splitting of NMR spectra for $^{51}$V allowed us to distinguish between spectra from transverse chains of V as a function of temperature. Our data revealed that (1) the MT is not abrupt, but rather there is a microscopic coexistence of pre-transformed cubic phase and transformed tetragonal phase over a few K below and above $T_m$, while (2) no pre-transformed phase can be found at $T_c$, and (3) the Martensitic lengthening of one axis occurs predominantly in a plane perpendicular to the crystal growth axis, as twinned domains.

1. **Introduction**

The Martensitic transformation (MT) is a second-order displacive structural transformation from cubic to tetragonal symmetry that has long been observed in A15 superconductors $Nb_3Sn$ and $V_3Si$, at a temperature $T_m$ a few Kelvins above the superconducting critical temperature $T_c$. Though known since the 1960s [1-7], its connection to the transition to the superconducting phase continues to be an open question of great interest today [8-12] – as is the connection between superconductivity and structural transformations in general, e.g., the charge-density-wave phase transition in the normal state of $NbSe_2$ and the orthorhombic-tetragonal transition in YBCO, and more recently, its potential application in quantum information technology [13]. $V_3Si$ in its normal, non-superconducting and pre-transformed state is cubic with a lattice constant a = 4.718 Å, where the unit cell consists of a body-centered arrangement of Si atoms, plus a pair of V atoms on each face so that the V atoms form orthogonal "chains" along all three principal axes, as shown in Figure 1. The transformation to tetragonal phase is characterized by a 0.15% lengthening along one of the three orthogonal directions – to lattice constant c = 4.725 Å – while the other two axes shrink by 0.06% to lattice constant a' = 4.715, such that c/a' = 1.0025.

The MT is so subtle that the effects can be discerned only in single-crystal samples of high purity, where the effects can be significant. The sample in this study is a single crystal grown by floating-zone method, as described by Christen *et al* [14] and designated as "MP3". In this sample, the MT has manifested as a second-order discontinuity in the temperature dependence of specific heat, a "kink" in temperature dependence of susceptibility, and an inflection "bump" in resistivity at temperature $T_m$ – the latter of which is shown in Figure 2. MP3 has a high residual resistivity (RRR) of 47, shown in 2(a). The field dependence of $T_c$ and that of $T_m$ are shown in 2(b). The same measurements performed on a sample of less purity, with RRR of 7, revealed none of these features; see Fig. 2(c). The virtual non-existence of impurity phases in MP3 and in the ingot from which it had been sliced has also been confirmed by XRD [14]. This prompted several studies on this and other samples from the same ingot from 1983 to the present that would have otherwise been impossible – e.g., observation of free flux flow using transport measurements [15] and a square-triangular phase transformation in the vortex lattice as observed by small-angle neutron scattering [16], as well as a study of crystal lattice dynamics through inelastic x-ray scattering [17]. The current study subjects this same sample to the sensitive probe afforded by pulsed Nuclear Magnetic Resonance (NMR) using [51]V nuclei. The [51]V isotope is an ideal probe for studying structural transformations in this material because its non-spherical nucleus interacts with the electric field in the crystal lattice, modifying the nuclear energy levels resulting in a quadrupolar-split NMR spectrum.

While $V_3Si$ has been studied by NMR in the past, previous work has either not prioritized the MT, utilized lower fields[18-22], and/or been conducted on powder samples [9,11]. Consequently, we were able to monitor the MT in unprecedented detail, affording clear evidence of the co-existence of transformed and pre-transformed phases over a finite range of temperatures above $T_c$, and the non-existence of pre-transformed phase below $T_c$. We also find that the lengthening of axes seems to occur only perpendicular to the crystal growth axis, likely due to strain.

Under high magnetic fields, if the *local* symmetry of the nuclear probe of spin $I > 1/2$ is lower than cubic, the nuclear quadrupole moment $Q$ interacts with the electric field gradient (EFG) of the surrounding charges. This perturbs the otherwise equally spaced nuclear Zeeman levels, which, up to the first order [23], are given by:

$$E_m = -\gamma\hbar H m + \frac{1}{6}h\nu_Q \left(\frac{3\cos^2\theta - 1}{2}\right)[3m^2 - I(I+1)]$$

where $m$ is the magnetic quantum number and $\theta$ is the angle between the principal axis of largest EFG and the applied field $H$. The quadrupole splitting parameter $\nu_Q$ is related to the EFG, $eq$ as

$$\nu_Q = \frac{3e^2qQ}{2hI(2I-1)}$$

heretofore called *quadrupole frequency*. A typical NMR spectrum consists of equally spaced $2I$ transitions whose spacing is dependent only on $\nu_Q$ and $\theta$. The quadrupole frequency is conveniently measured as the largest spacing between these transitions which occurs when $\theta = 0$, *i.e*, when the principal axis of symmetry of the EFG is aligned parallel to $H$. It is this sensitivity of the NMR spectra to crystal EFG and thus to the position and symmetry of atoms in the unit cell that makes this technique a powerful tool to study the MT. The [51]V nuclei have spin $I = 7/2$, gyromagnetic ratio $\gamma/2\pi = 11.193$ MHz/T and quadrupole moment $Q = -0.052$ barns. Due to the chain configuration, the local symmetry of the vanadium atoms in $V_3Si$ is tetragonal (both in normal and martensitic phases), with the largest EFG pointing along the V chains. The high-temperature $\theta = 0$ spectrum for a single V site consists of seven lines corresponding to $\Delta m = \pm 1$ transitions, each separated by $\nu_Q$. However, the NMR spectra in V$_3$Si is further complicated by signals from two other orthogonal sets of V chains running perpendicular to $\mathbf{H}$, $\theta = 90°$, producing another set of seven lines, but separated instead by $\nu_Q/2$. The spectra measured at $T =$ 90 K was confirmed to show all these features, as are shown in Figure 3, where arrows identify $\theta = 0$ and $90°$ spectral lines. Note that the greater intensity of the spectrum for V chains along $\theta = 90°$ is consistent with the fact that there are two chains along that direction for every chain along $\theta = 0$.[24] This level of detail, not possible in a powder sample, is what allows for a direct and more accurate measurement of quadrupole splitting – which in turn would be sensitive to any changes in the crystal structure, as the temperature is changed across $T_m$. While this splitting is known as $\nu_Q$ in the pre-transformed cubic phase, it is unknown in the transformed, tetragonal phase: it is this post-MT splitting, later designated as $\nu_{\alpha\alpha}$, that we investigate in this study.

The NMR experiments were performed at the National High Magnetic Field Laboratory using the home-made MAGRes2000 spectrometer with quadrature phase-sensitive detection. Spectra were obtained by fast fourier-transform of the spin-echo following a ($\pi/2$– $\pi$) pulsed sequence while varying the temperature in a magnetic field that is uniform to within 10 ppm/cm DSV, in a superconducting magnet with a range of up to 11 T. The sample is an 8-mm-long cubic prism of width ~1 mm, mounted in a copper pickup coil customized for optimal resonance signal – i.e., with long axis [010] coinciding with the rotational axis and parallel to the coil axis– as also shown in the inset of Figure 3. Sample alignment was made possible by a goniometer stage with a resolution of 0.10°, and optimized such that maximal spacing between spectral lines is achieved. The field of $H = 6.67$ T was chosen so as to minimize any overlap with the resonance frequency of $^{63}$Cu (from the NMR coil used as a reference, also shown), which placed the resonance frequency $\nu_o$ at 75.45 MHz. At this field, MP3 has $T_c$ (6.67 T) = 12.8 K and $T_m$ (6.67 T) = 21.1 K, also shown in Figure 2(b). Above this field, the spectra start to broaden substantially which limits our ability to resolve the NMR peaks at low temperatures.

## 2. Results

Identifying and tracking the spacing of peaks in Figure 3 through temperature and phase change was critical at lower temperatures. Peak fittings are further complicated by the spectral overlap due to broadening and Knight shift due to superconductivity. On the other hand, the fact that each set of quadrupolar spectra is collectively defined puts constraints on the fit parameters and also allows identification of any peaks that may be hidden and/or unresolved. The challenge of monitoring the temperature dependence of the spectra was found to be mitigated by visualizing them as a three-dimensional contour plot in lieu of the customary "waterfall" plot, with decreasing temperature along the y axis and signal intensity along the z. Examples of this are shown in Figure 4. Figure 4(a) shows spectra in the cubic, pre-transformed phase, at $T = 90$ K, where vertical dashed lines mark the peaks for $\theta = 0$ (∥H), while the black triangles at the bottom indicate peaks for $\theta = 90º$ (⊥H). Here, we confirm that the peak spacings are constant as a function of temperature for both orientations, and the spacing for ⊥H is half of $v_Q$, the spacing for ∥H. (Note, by contrast, the temperature independence of the spectral line for Cu, marked by the red arrow.) Also apparent here is the behavior of the Knight shift K. As $T$ drops, the shift for ⊥H spectra is significantly more pronounced than that for ∥H. This data is consistent with the isotropic Knight shift, $K_{iso} = (2K_\perp + K_\parallel)/3$, previously measured by Clogston et al.[7] in powder samples. Figure 4(b) shows contours at temperatures in the vicinity of the MT, where dramatic changes are readily seen. As temperature crosses $T_m$, one can discern cubic-phase spectral lines ceasing to exist (shown by blue arrows) while tetragonal-phase lines appear (red arrows). For the latter, there is also a drastic frequency shift below $T_c$, which is expected from the formation of Cooper pairs in the superconducting phase.

To account for the quadrupole splitting for all sites, especially in the transformed, tetragonal phase, a more detailed modeling of the EFG was necessary. We performed point-charge calculations in order to help find consistency in assigning the sites to their respective spectra. As detailed in the Appendix, the EFG tensor was recast into a quadrupolar splitting tensor, $v_{\alpha\alpha}, \alpha = x, y, z$, yielding values that were found to be consistent with observed spectra. From this analysis, we have considered three possible domains for the tetragonal phase; an exaggerated depiction is shown in Figure 5(a), with each unit cell of domains labelled P, Q, and R relative to the direction of the magnetic field. Each "dumbbell" represents the V chain on each face of the unit cell and the label indicates the relevant splitting parameter expected to result from these sites if the magnetic field is oriented in the vertical direction as shown. We designate those atoms along elongated chains as "*c*" and those atoms along contracted chains as "*a*". We use the subscripts ∥ and ⊥ to refer to the orientation of the associated chain parallel or perpendicular, respectively, with respect to the applied field.

Figure 6 summarizes the main results of this study. Here, plotted is the temperature dependence of the quadrupole splitting $\nu_{\alpha\alpha}$ for *each set of seven lines* extracted from the spectra. There are at most five sets that were unambiguously resolved, three of which survives below 17K. The horizontal axis is logarithmic to magnify the MT region between $T_c$ (H=6.7 T) = 12.8 K (green line) and $T_m$(H=6.7T) = 21.1 K (purple line). Data for the cubic phase are marked by light-blue bars, one each for ∥H and ⊥H , where $\nu_Q$ = 206 kHz. For the transformed, tetragonal phases, the ⊥H data are marked by magenta bars, and ∥H data by yellow bars. Note that each of these sets confirms that the values of quadrupole splittings changes upon undergoing structural transition but remain independent of temperature at each phase nonetheless. These plots reveal the following noteworthy features:

First, spectra for the cubic, pre-transformation phase were found to *persist* below $T_m$, while spectra for the transformed, tetragonal phases were found *above* $T_m$. In Fig. 6, we see that the cubic (blue) phase extends down to 4 K below $T_m$ , and the tetragonal (yellow and magenta) phase extends up to at least 4 K above $T_m$, highlighting a zone of co-existence 4 K above and below $T_m$ , marked by the green bar.

Second, no conclusive evidence of spectra for cubic phase could be discerned below the zero-field $T_c$ of 17 K. This is consistent with the notion that the transformation is completed by the time superconductivity is reached, implying a close connection between MT and the superconducting transition.

Third, our calculations (see Appendix) revealed an absence of spectra in our data corresponding to the P domain, which would have been characterized by splittings corresponding to c∥ and $a_y$ in Figure 5(a) (i.e., $\nu_{\alpha\alpha}$ splittings for sites corresponding to α = z for the c-site and α = y for the a-site). This strongly implies that the tetragonal lengthening of V chains occurs predominantly along the plane (Q and R domains). This is consistent with a proposition by Buttet et al.[19], that the tetragonal lengthening ("*c*") occurs predominantly within the plane perpendicular to the crystal's growth axis. We confirmed that the growth axis of our sample is perpendicular to the long axis and therefore parallel to the field *H*. To test this consistency, it would be interesting to repeat a similar experiment on single crystals with various levels of strain, as set up by different post-annealing conditions.

## 3. Conclusions

This study has shown that single-crystal NMR spectroscopy allows us to see manifestations of the Martensitic transformation (MT) in unprecedented detail, by distinguishing microscopically between orthogonal domains. This afforded strong evidence that: (1) the MT lengthening occurs predominantly in one plane, perpendicular to direction of strain from crystal growth; (2) all transformation to tetragonal phase is essentially completed by the time the sample has transitioned into superconducting state; and (3) the MT occurs more gradually in the crystal, with a phase of coexistence of both transformed tetragonal phase and pre-transformed cubic phase over a significant range of temperatures, at least 4 K above and below the MT temperature $T_m$ detected by resistivity and susceptibility measurements.

## 4. Acknowledgments

AAG is grateful to D. K. Christen (retired, Oak Ridge National Laboratory [ORNL]) and J. R. Thompson (retired, University of Tennessee Knoxville), for insightful discussions and for archival data and information on $V_3Si$ samples, especially MP3; and S. Moraes and O. Gafarov, who assisted in the original transport resistivity measurements and analysis of NMR data while undergraduate researchers at the University of South Alabama. Past work on this project was funded by the University of South Alabama, the National Science Foundation (NSF), and the Research Corporation. Past work at ORNL was funded by the US Department of Energy. AAG is grateful to the National High Magnetic Field Laboratory for hosting him during his Sabbatical from the University of South Alabama. Work at the National High Magnetic Field Laboratory is supported by NSF Cooperative Agreement Numbers DMR-115749 and DMR-1644779, and by the State of Florida.

## 5. Appendix: EFG calculations and site assignment

The EFG is calculated according to:

$$q^i_{\alpha\beta} = \frac{\partial^2 V(i)}{\partial r^\alpha \partial r^\beta} = \sum_{j \neq i} e_j \frac{3r^\alpha_{ij} r^\beta_{ij} - r^2_{ij}\delta_{\alpha\beta}}{r^5_{ij}}$$

where $V(i)$ is the potential at the $i^{th}$ atom; $\alpha, \beta = x, y, z$; $r^\alpha_{ij}$ is the $\alpha^{th}$ component of the vector distance $\boldsymbol{r}_{ij}$ from $i^{th}$ to $j^{th}$ atom; $e_j$ is the charge of the $j^{th}$ atom; $r_{ij} = |\boldsymbol{r}_{ij}|$ and $\delta_{\alpha\beta}$ is the Kronecker delta function. The sum is over all the lattice coordinates $j$. Since the matrix is diagonal in the principal axes of symmetry which coincide with the V₃Si crystal axes, then $\alpha = \beta$. Generally, we can recast the principal components of the EFG tensor in terms of quadrupole splitting tensor. In frequency units

$$\nu_{\alpha\alpha} = \frac{3eq_{\alpha\alpha}Q}{2hI(2I-1)}.$$

Choosing $\nu_{zz} \geq \nu_{xx} \geq \nu_{yy}$ we take the quadrupole frequency $\nu_Q$ as the largest component of the EFG, $\nu_Q = |\nu_{zz}|$, as described in the main text, and define the asymmetry parameter $\eta = \frac{\nu_{xx} - \nu_{yy}}{\nu_{zz}}$. Also, from Laplace equation: $\nu_{xx} + \nu_{yy} + \nu_{zz} = 0$. In the cubic phase, all the V sites are equivalent. In contrast, the tetragonal phase has two site-differentiated V atoms: (1) the *c*-sites form the chain along the elongated axis (designated as tetragonal crystal *c*-axis), and (2) the *a*-sites, form the chains that run along the "squeezed" *a*- and *b*-crystal axes. Note that in both *normal cubic* and *tetragonal* phases, the *local* symmetry of the V atoms is (or close to) tetragonal, $\eta \cong 0$. The above lattice sum is performed over the radius of 50 Å using charges +1.8 for Si and -0.6 for V [25] consistent with the best estimate electron transfer from Si to V according to x-ray data [25]. The calculated EFG is corrected by a factor of $(1 - \gamma_\infty) = 8$ from the antishielding term [26]. In Table 1 we show the calculated quadrupolar splittings (in *kHz*) for different sites/phases in V₃Si. There is a very good agreement between the calculated and experimental values. Small discrepancies are likely due to uncertainties in the antishielding term which maybe different at different sites. Irrespective of the actual magnitudes, calculations show that as the system undergoes MT there is a *decrease* in the EFG in the elongated chains (site *c*) and an *increase* in the squeezed chains (site *a*). As will be argued later, we note on the table that $v_{yy}$ for the *a*- and $v_{zz}$ for *c*- tetragonal sites were not directly observed in the experiment. Their tabular values were calculated based on other parameters on the table and using the Laplace equation.

Table 1. Result of Lattice sum calculation of EFG on the $^{51}$V sites in $V_3Si$.

| Phase-site | $v_{xx}$ | | $v_{yy}$ | | $v_{zz}$ | | $\eta$ | |
|---|---|---|---|---|---|---|---|---|
| | Calc. | Expt. | Calc. | Expt. | Calc. | Expt. | Calc. | Expt. |
| Cubic-*c* | 99.35 | 103 | 99.35 | 103 | -198.7 | 206 | 0 | 0 |
| tetragonal-*a* | 100.8 | 121(1) | 98.56 | *118 | -199.3 | 216(3) | 0.0113 | * |
| tetragonal-*c* | 98.68 | 93(2) | 98.68 | 93(2) | -196.7 | *186 | 0 | 0 |

* not directly observed.

*Arguments for the absence of P domain:*

Is there a preferential direction for which the MT occurs? We attempt to answer this question by determining the origin of spectral lines in the tetragonal phase. In Fig. 5(b) we show the eigenvectors of the EFG calculated for each V site in the tetragonal unit cell, which corresponds to the principal axes of $v_{\alpha\alpha}$. In order to simplify notation we denote for the *a*-sites: $a_{||} = v_{zz}$ and $a_\alpha = v_{\alpha\alpha}$, for $\alpha = x, y$. Correspondingly, for the *c*-sites: $c_{||} = v_{zz}$ and $c_\perp = v_{\alpha\alpha}$, for $\alpha = x, y$, since $\eta = 0$ for this site. As a consequence, if the crystal is oriented such that the elongated axis is parallel to the magnetic field direction, the NMR signals would originate from sites containing the $c_{||}$ and two $a_y$ splittings. If the magnetic field is applied perpendicular to the elongated axis (either *x* or *y* in Fig. 5(b)), NMR will pick up $c_\perp$, $a_{||}$ and $a_x$ corresponding sites. For the cubic phase, there is only a single site which has the same symmetry as the *c* site.

We start with the assumption that there is an equal probability of unit cell elongation along all three crystal axis directions, as described in Figure 5(a) (see main text). Referring back to Fig. 6, there are five sets of quadrupolar splittings, $v_{\alpha\alpha}$ extracted from the spectra: 216, 206, 121, 103 and 93 kHz. In identifying the site and phase to which of each these belongs, we argue the following:

1. The 206-kHz set unambiguously belong to the cubic ($c_{||}$) phase as it is the only phase that exists at high temperatures. The signals from the 103-kHz set originate from the same cubic site but with chains perpendicular to the field and are thus assigned to cubic $c_\perp$. These two sets are consistent with having asymmetry parameter $\eta = 0$, as expected from Table 1.

2. We assign the 216 kHz set to $a_{||}$ below $T_m$ in the tetragonal phase. These are the sites on the contracted chain where the chain is parallel with the field. This assignment is consistent with the calculations (Table 1) that show that the EFG of the tetragonal $a$ sites is larger than its cubic value.

3. The assignments for the two remaining splittings, 121 kHz and 93 kHz, need careful consideration. There are four lines that are still unaccounted for, all in tetragonal phase: $c_{||}, c_\perp, a_y$ and $a_x$. There is also a possibility that some of these lines overlap. However, considering the relative intensities of these seven-line sets, which we found to be equal to within 10-15% using Gaussian fits, we can confidently conclude that each of these sets are defined by unique sites and there was no significant overlap. With this in mind, it is only reasonable that some of the four lines are unobservable. Using the process of elimination and guided by the theoretical calculation, we conclude the following:

a. Both $a$ and $c$ sites must be represented in the spectrum. Thus, one of the $c$ sites, $c_{||}$ or $c_\perp$, and one of the $a$ sites, $a_x$ or $a_y$, must be accounted for.

b. The 121-kHz and 93-kHz splittings, because their values are ≲ 50% of the cubic value, are more likely to originate from chains that are perpendicular to the field, i.e. $\theta = 90°$. This eliminates $c_{||}$. In addition, referring to Table 1, the 121-kHz splitting would be already too low compared to the cubic value had this been assigned to $c_{||}$. *We therefore conclude that $c_{||}$ is not observed.*

c. From calculations, the value of $c_{||}$ must be less than the cubic value of 216 kHz. This limits the value of $c_\perp < 108$ kHz. Thus, this cannot be 121 kHz and the only option is to take the 93-kHz splitting for $c_\perp$.

d. Using a similar argument, the smaller of the perpendicular components of the $a$ site, namely $a_y$, must be less than $a_{||}/2 = 108$ kHz. Thus, the remaining 121-kHz splitting is assigned to $a_x$. *We conclude that $a_y$ was not observed.*

All of the above assignments are mutually consistent and fit seamlessly to theoretical calculations. Using the calculated $\eta$, the $a$ site would have $a_y \sim 118$ kHz and based on our assigned value for $c_\perp$, $c_{||} \sim 186$ kHz, as we indicated in Table 1. It is worth to note that our site assignments differ with that proposed by Buttet et al.[19], who have also identified three sets of splittings in the tetragonal phase whose values are in very close agreement to ours. They assigned, in the order of EFG strength, $a_{||} \sim 212.5$ kHz, $c_\perp \sim 115.7$ kHz, and $a_\perp = a_{x,y} \sim 92.6$ kHz, *i.e.* their assignments for $c_\perp$ and $a_x$ are reversed from ours. They assumed that $c_{||}$ is greater than the cubic value and chose $c_\perp$ to be the higher of the two frequencies. The consequence of their assignment is that the resulting asymmetry parameter for the *a* site is an order of magnitude larger than what the calculations suggest. Our site assignment is more consistent with theory in terms of the relative magnitude of change in EFG during the transition for both sites *c* and *a* and is also more amenable to the value of $\eta$. Nevertheless, this difference in site assignments does not change the conclusions we reached in this work.

After all these considerations, we conclude that the sets of lines corresponding to $c_{||}$ and $a_y$ are absent in our spectra. This means that the P domain (Fig. 5(a)) may not exist and narrows the possibilities of chain elongation to the direction perpendicular to the field, *i.e.* domains Q and R. Our NMR data cannot distinguish between these twin domains.

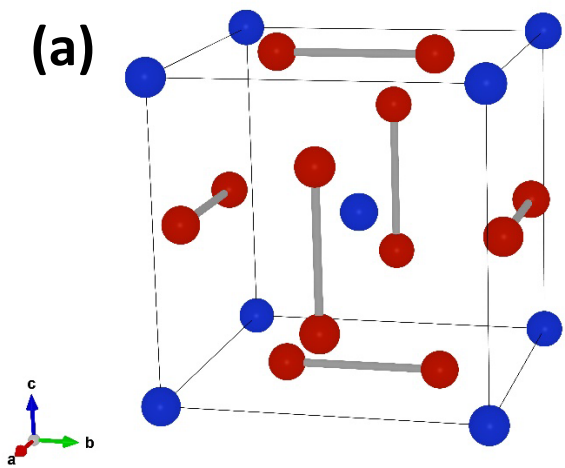 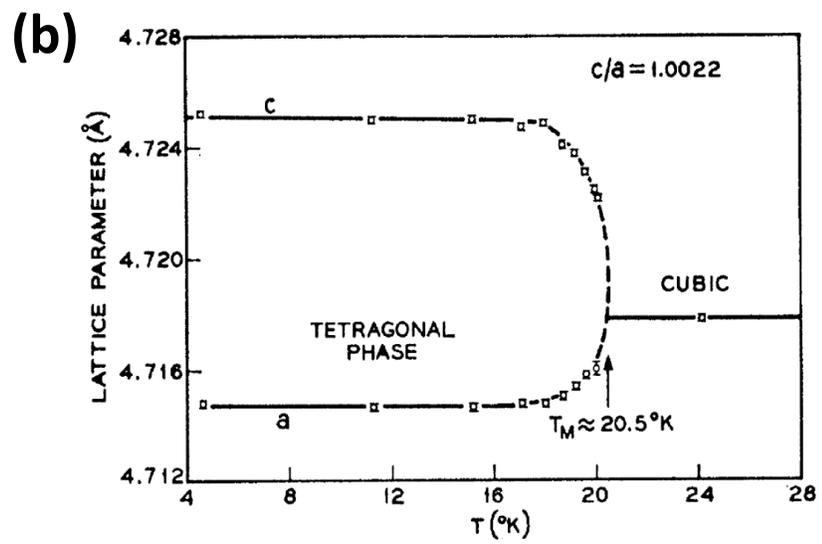

**Figure 1**

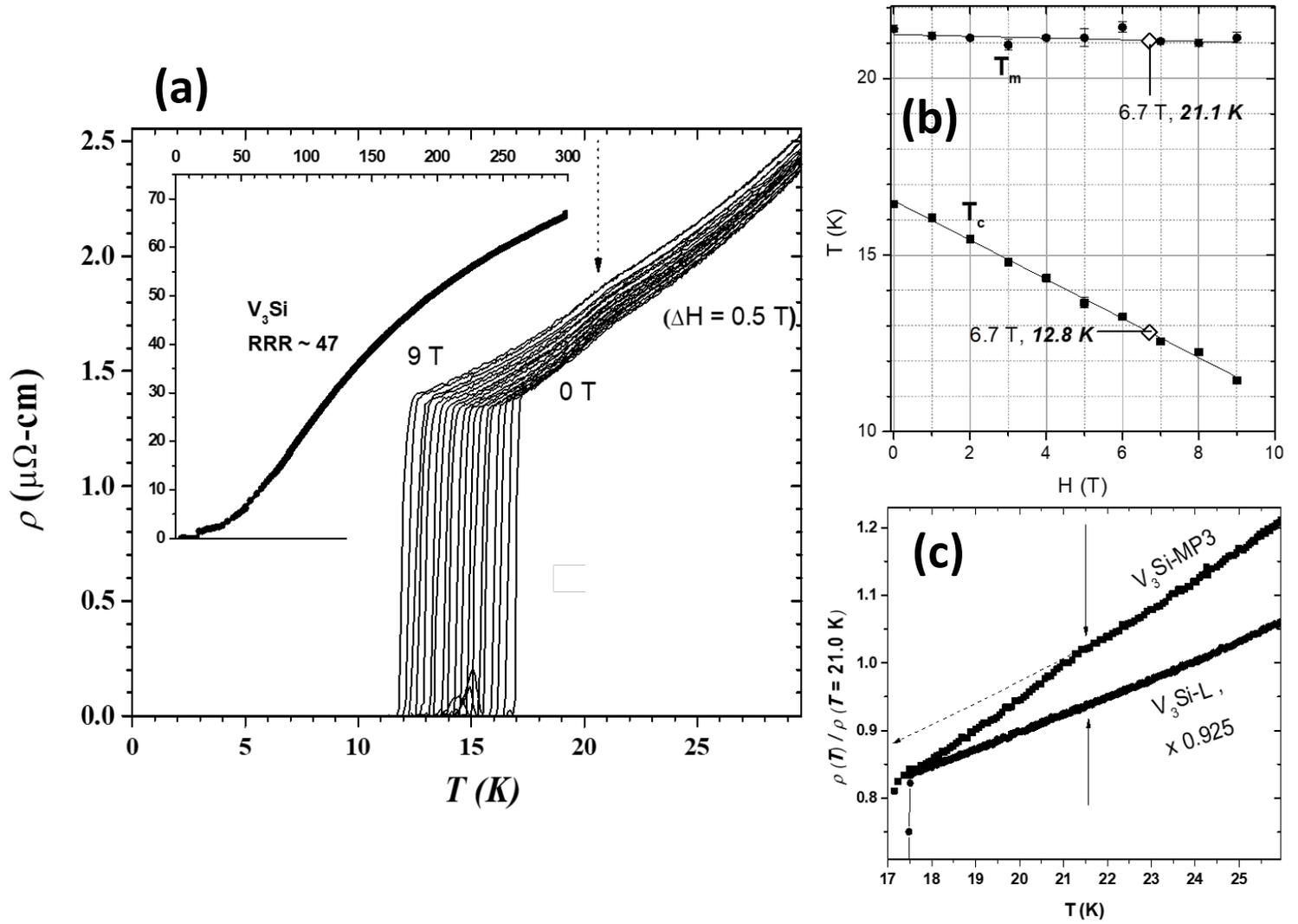

**Figure 2**

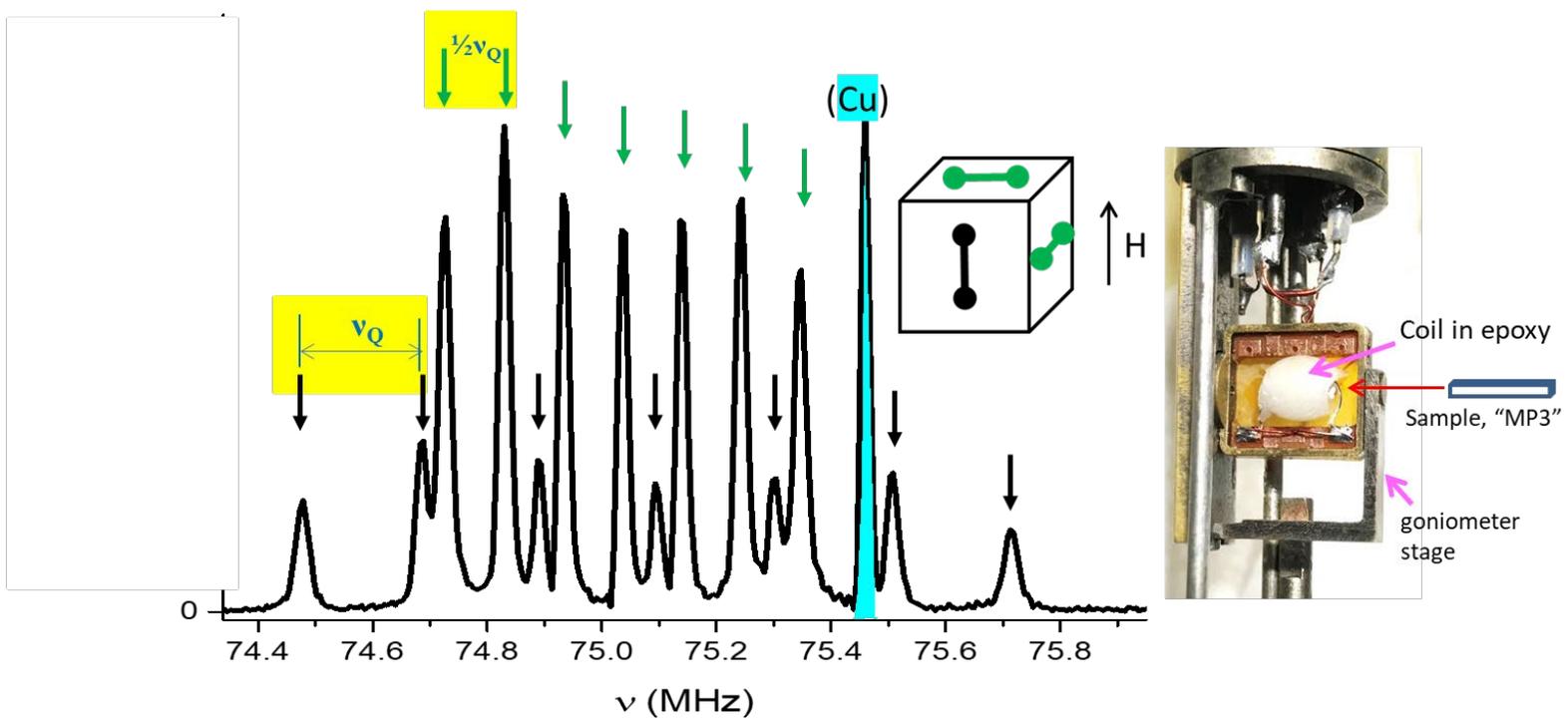

**Figure 3**

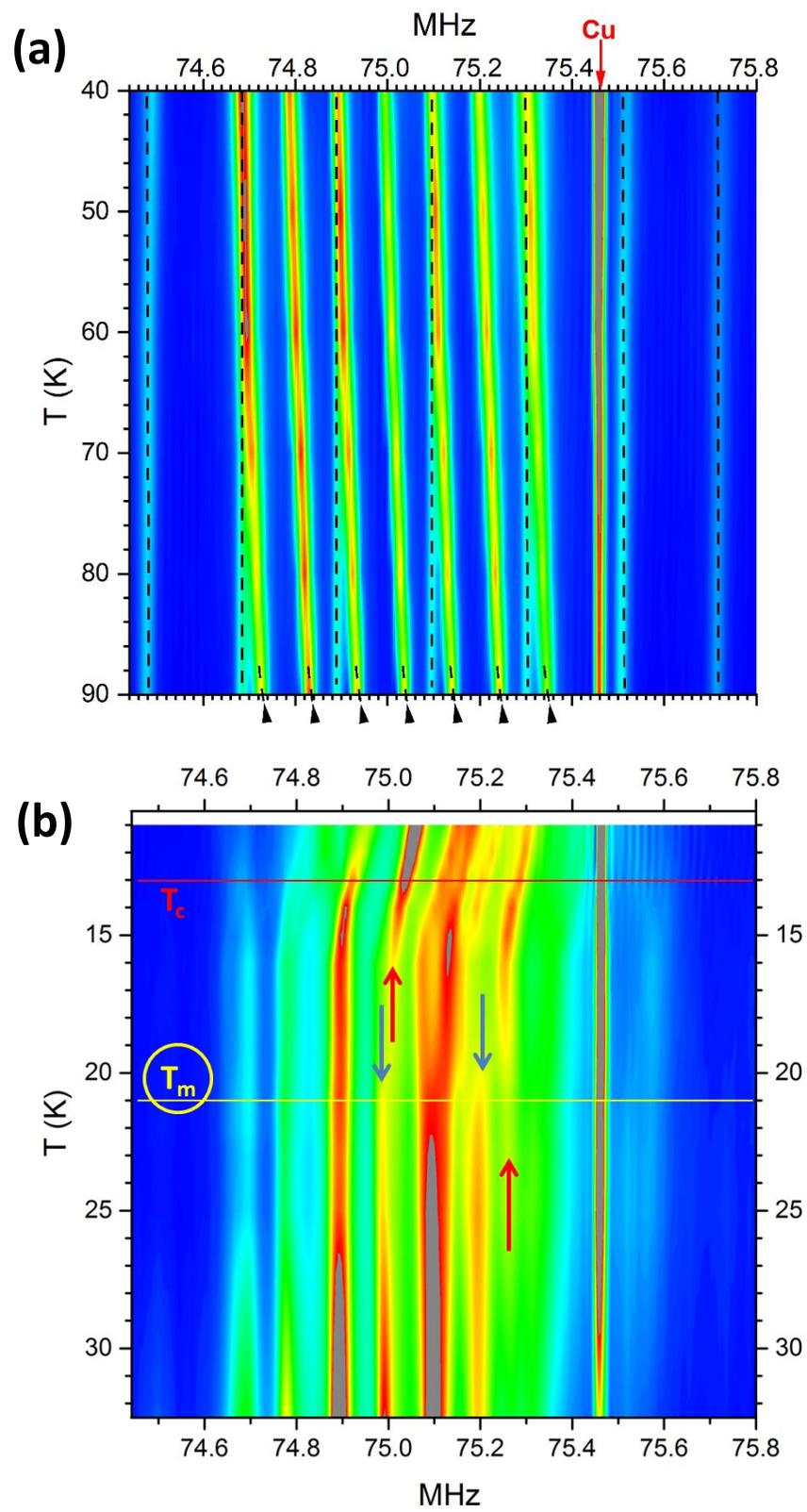

**Figure 4**

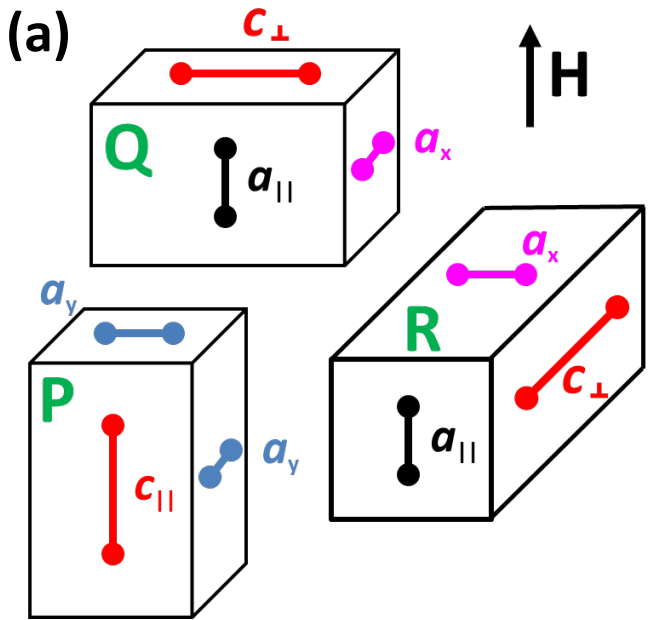 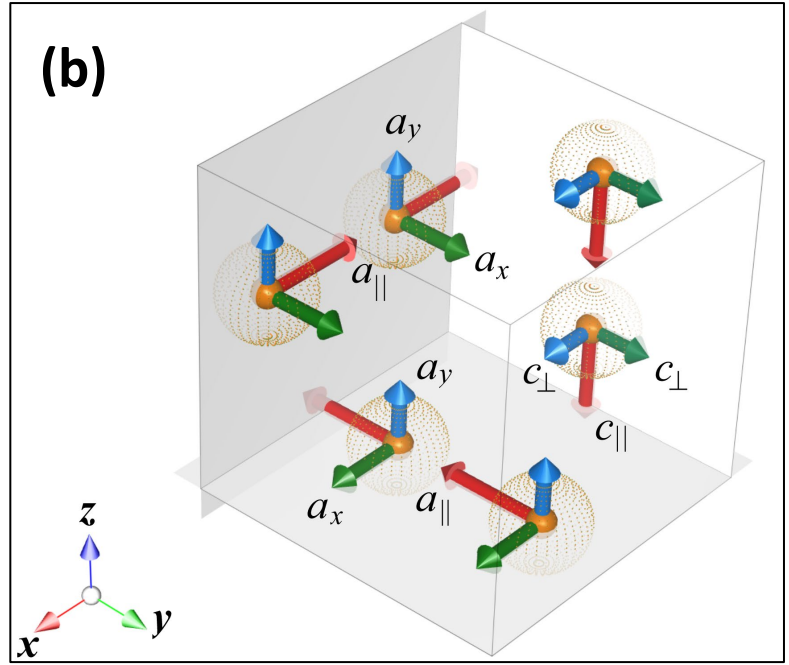

**Figure 5**

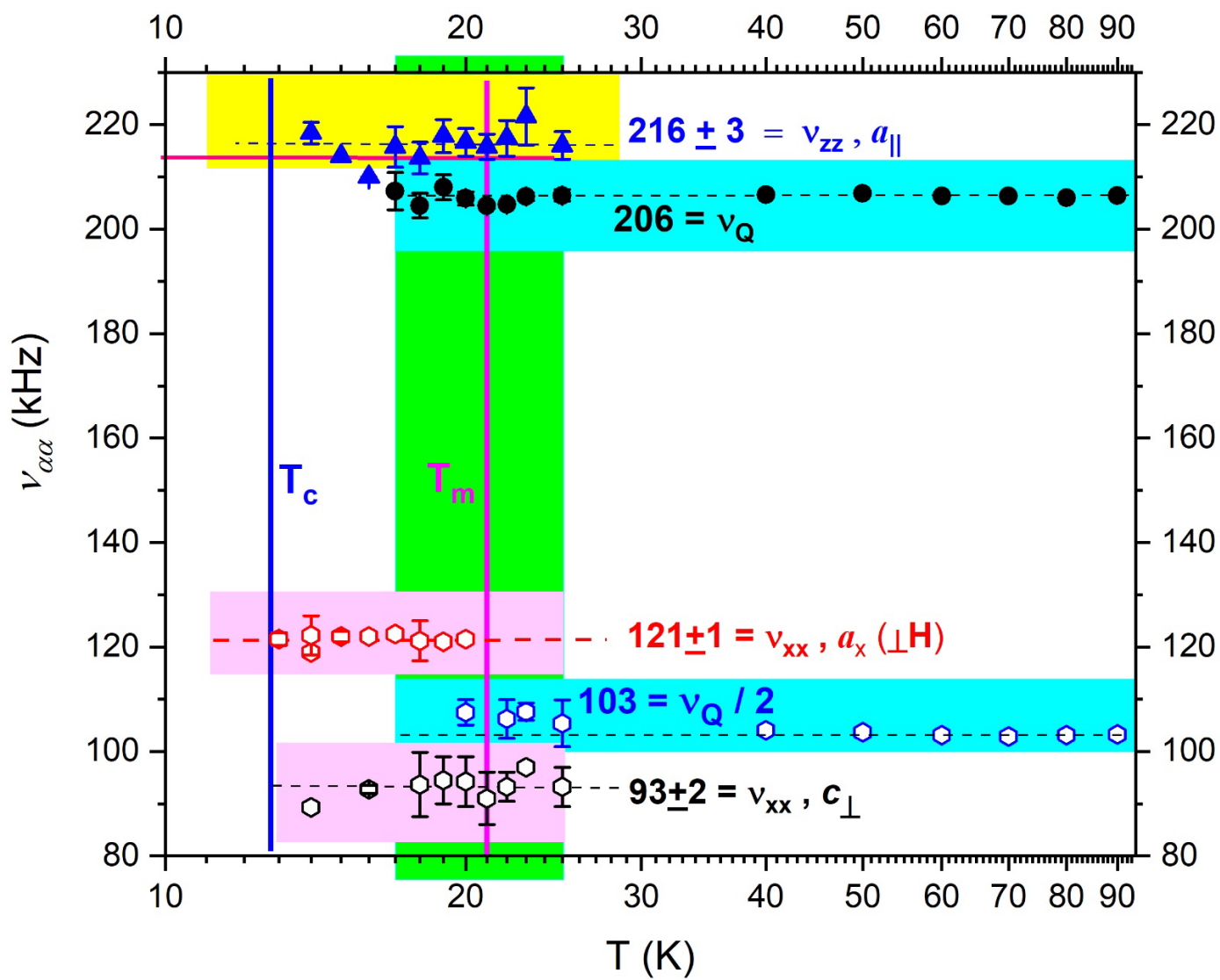

**Figure 6**

**Figure Captions:**

**Figure 1: (a)** Cubic structure of $V_3Si$ in the non-superconducting state, showing a body-centered arrangement of Si (blue), with chains of V (red) arranged along the faces in all three orthogonal directions. **(b)** Original plot in the paper by Batterman and Barrett, Ref. 1, illustrating the MT as first observed in 1961 by x-ray diffractometry.

**Figure 2: (a)** Resistivity vs. temperature plot at magnetic fields up to 9 T, for sample "MP3" with a high residual resistivity ratio (RRR) of 47 (as shown over more temperatures in the inset), showing the drop in $T_c$ and the near-field-independence of the inflection "bump" marking the MT at temperature $T_m$ . **(b)** summarizes these field dependences, also indicating $T_c$ and $T_m$ at the field 6.7 T chosen for the study (see text). **(c)** shows a sample with RRR ~ 7 showing no sign of the MT bump at $T_m$, as contrasted with MP3. -

**Figure 3:** Quadrupole spectra at T = 90 K for chains perpendicular to H (green arrows, green chains in the cartoon) and parallel to H (black arrows, black chain in the cartoon), showing the expected spacings of $\frac{\nu_Q}{2}$ and $\nu_Q$. The peak for $^{63}Cu$ is marked in light blue, with minimal overlap at field, H = 6.7 T. Inset shows the sample coil set on a goniometer stagewith the long axis of the sample along the rotation axis and perpendicular to H.

**Figure 4:** Three-dimensional contour plots of V spectra, with temperature on the ordinate and colored contours indicating intensity **(a)** In the pre-transformed, cubic phase, spectral peaks are marked by vertical dashed lines for chains parallel to field H, and by triangles at bottom for chains perpendicular to H. The temperature-invariant peak for copper is also marked. **(b)** Spectra as temperature decreases in the vicinity of MT: Between temperatures $T_m$ (21 K) and $T_c$ (13 K) are examples of the discontinuation of cubic phase, blue arrows, and the emergence of tetragonal phase, red arrows.

**Figure 5:** Results from the determination of observed quadrupolar splittings (from Appendix): **(a)** Cartoon of the possible domains, P, Q, R of tetragonal phase, with V chains lengthening in one of three possible orthogonal directions relative to ***H***, see text. **(b)** Calculated principal axes (eigenvectors) of the local EFG tensor in the tetragonal phase, with elongation direction along

the z-axis. The largest EFG component has its eigenvector shown in red, then green, and blue the smallest. See Appendix for notation.

**Figure 6:** Summary of spectral line spacings for all designated V chains discerned in the study, see text and Figure 5 for designations. In the tetragonal phase, yellow bars indicate the temperature extent at which spectra for ∥H chains were found, with pink bars marking the temperature extent for ⊥H chains. The light blue bars show the extent for chains associated with the cubic phase (both ∥H and ⊥H). The green bar shows the regime of co-existence of transformed and pre-transformed phases.